\documentclass[aps,prd,twocolumn,groupedaddress]{revtex4}
\usepackage{graphicx}
\usepackage{subfigure}
\usepackage{dcolumn}
\usepackage{bm}
\usepackage{color}
\usepackage{longtable}
\usepackage{supertabular}
\usepackage[T1]{fontenc}
\usepackage{epstopdf}
\usepackage{amsmath}
\usepackage{setspace}
\usepackage{float}
\usepackage{textcomp}
\usepackage{indentfirst}

\makeatletter

\newcommand{\Rmnum}[1]{\expandafter\@slowromancap\romannumeral #1@}
\makeatother

\begin{document}
\title{Arm locking using laser frequency comb}
\author{Hanzhong Wu$^{1,a}$}
\author{Jun Ke$^{2,a}$}
\author{Panpan Wang$^{1}$}
\author{Yu-Jie Tan$^{1}$}
\author{Dian-Hong Wang$^{2}$}
\author{Jie Luo$^{2}$}\email[E-mail: ]{luojiethanks@126.com}
\author{Cheng-Gang Shao$^{1}$}\email[E-mail: ]{cgshao@hust.edu.cn}

\affiliation{
$^{1}$ MOE Key Laboratory of Fundamental Physical Quantities Measurement, Hubei Key Laboratory of Gravitation and Quantum Physics, PGMF and School of Physics, Huazhong University of Science and Technology, Wuhan 430074,  People's Republic of China\\
$^{2}$ School of Mechanical Engineering and Electronic Information, China University of Geosciences, Wuhan 430074, People's Republic of China\\
$^{a}$ These authors contribute to this work equally
}

\date{\today}
\begin{abstract}
The space-borne gravitational wave (GW) detectors, e.g., LISA, TianQin, TaiJi, will open the window in the low-frequency regime (0.1 mHz to 1 Hz) to study the highly energetic cosmic events, such as coalescences and mergers of binary black holes and neutron stars. For the sake of successful observatory of GWs, the required strain sensitivity of the detector is approximately ${10^{-21} \rm /Hz^{1/2}}$ in the science band, 7 orders of magnitude better than the state of the art of the ultra-stable laser. Arm locking is therefore proposed to reduce the laser phase noise by a few orders of magnitude to relax the burden of time delay interferometry. During the past two decades, various schemes have been demonstrated by using single or dual arms between the spacecraft, with consideration of the gain, the nulls in the science band, and the frequency pulling characteristics, etc. In this work, we describe an updated version of single arm locking, and the noise amplification due to the nulls can be flexibly restricted with the help of optical frequency comb. We show that, the laser phase noise can be divided by a specific factor with optical frequency comb as the bridge. The analytical results indicate that, the peaks in the science band have been greatly reduced. The performance of the noise suppression shows that the total noise after arm locking can well satisfy the requirement of time delay interferometry, even with the free-running laser source. When the laser source is pre-stabilized to a Fabry-Perot cavity or a Mach-Zehnder interferometer, the noise can reach the floor determined by the clock noise, the spacecraft motion, and the shot noise. We also estimate the frequency pulling characteristics of the updated single arm locking, and the results suggest that the pulling rate can be tolerated, without the risk of mode hopping. Arm locking will be a valuable solution for the noise reduction in the space-borne GW detectors. We demonstrate that, with the precise control of the returned laser phase noise, the noise amplification in the science band can be efficiently suppressed based on the updated single arm locking. Not only our method allows the suppression of the peaks, the high gain, low pulling rate, it can also serve for full year, without the potential risk of locking failure due to the arm length mismatch. We finally discuss the unified demonstration of the updated single arm locking, where both the local and the returned laser phase noises can be tuned to generate the expected arm-locking sensor actually. Our work could provide a powerful method for the arm locking in the future space-borne GW detectors.
\end{abstract}

\maketitle


\section{INTRODUCTION}\label{section1}
In 2016, the Laser Interferometer Gravitational-Wave Observatory (LIGO) announced the first detection of gravitational wave, predicted by Einstein's theory of general relativity [1-2], which was then honoured by the Nobel Prize in 2017. The direct observation of GWs provides a pathway to study the behaviors of massive celestial bodies, such as coalescences and mergers of binary black holes and neutron stars, supernovae explosions, and gamma-ray bursts, which is of great importance for the fields of fundamental physics and cosmology. Analogous to the electro-magnetic spectrum, GWs also cover a very wide frequency range. The ground-based GW detectors, such as LIGO, are sensitive to the GWs in the range from tens of Hz to several kHz, limited by the baseline length, the Earth's seismic, and gravity-gradient noises. Space-borne GW detectors are able to open the low-frequency window. National Aeronautics and Space Administration (NASA) and European Space Agency (ESA) proposed a joint mission of Laser Interferometer Space Antenna (LISA) [3], aiming to detect the GWs in the range of 0.1 mHz to 0.1 Hz. Similar space-based projects include DECIGO [4], TianQin [5], and TaiJi [6].

The space-borne GW detector is generally composed of a nearly equilateral triangle with an average baseline of ${10^8}$-${10^9}$ m. Each spacecraft houses two drag-free proof masses controlled by the capacitive sensors and micro-Newton thrusters, which act as the end points of the long arm. The acceleration noise of the proof mass itself should be well minimized via a disturbance reduction system, since it is actually a part of the sensitivity floor of the whole system. The GWs will modulate the relative length between the proof masses (generally pm level), and the length information can be detected by exchanging the coherent laser beams between the spacecraft. To maintain the signal-to-noise ratio, the optical transponder scheme is recommended, in which one laser works as the master laser while all the other lasers are offset phase locked to this master laser. The offset frequency is related to the Doppler shifts of the laser fields, normally 1 MHz per m/s for a 1064 nm laser source. Equivalently, the whole constellation shares one laser source. To detect the GWs, the strain sensitivity in the interested band should be better than ${10^{-21} \rm /Hz^{1/2}}$ , which is much better than the state of the art of ultra-stable lasers and ultra-stable oscillators. In the case of LIGO, the laser phase noise can be cancelled in real time with the equal-arm configuration of Michelson interferometer, leading to an excellent noise floor. However, for LISA and TianQin, it is not an easy work to fly an equal-arm interferometer in space. One technique to reduce the laser phase noise will be arm locking [7], where the laser frequency is synchronized to the arm length of the constellation. Since 2003, various schemes of arm locking have been proposed both theoretically [8-13] and experimentally [14-18], to reduce the potential risk of time delay interferometry (TDI). At the very beginning, scientists show that single arm locking is useful to reduce the laser phase noise in spite of the time-of-flight delay because of the long arm [7]. However, the first null locates in the science band (0.1 mHz - 1 Hz), leading to a significant noise amplification. Subsequently, dual arm locking has been designed exploiting two arms, to efficiently remove the first null out of the science band [10]. In addition, the start-up transients and the noise suppression at low frequencies can be well improved. To further improve the performance, modified dual arm locking has been finally presented, which is a smooth combination of common and dual arm-locking sensors [11]. The modification of dual arm locking allows the high gain of dual arm locking, and preserves the frequency pulling characteristics and low-frequency noise coupling of common arm locking. In this case, the clock noise (0.1 mHz to 20 mHz) and the spacecraft motion (20 mHz to 1 Hz) together contribute to the final noise floor. This level of performance can well satisfy the capability of time delay interferometry. However, complex design of the arm-locking controller is required, and the noise suppression performance will be highly affected by the arm length mismatch. Even, arm locking could be invalid when the arm length mismatch equals to zero. We try to modify the configuration of single arm locking, and hope to obtain a robust method with high gain, no nulls in the science band, low frequency pulling, and ease to use.

Optical frequency combs, as a powerful tool, have seen a vast number of applications, such as optical frequency measurement [19], optical communication [20], absolute distance measurement [21,22], frequency division [23], and precision spectroscopy [24], etc. These combs feature a series of precisely spaced, narrow emission lines in frequency space, and are with pulsed nature in the time domain. The two parameters, the repetition frequency and the carrier-envelope-offset frequency, can be tightly synchronized to an external clock. Consequently, all the frequency markers will share the same frequency precision as the clock [25]. In the converse manner, a microwave frequency can be also generated via frequency comb from optical frequency [26]. Consequently, this microwave frequency actually tracks the optical frequency via a down-conversion factor with high coherence. In this way, time delay interferometry with optical frequency combs has been demonstrated recently, which can simultaneously reduce the laser phase noise and the clock noise by a single step of TDI [27,28]. Yet, the deployment of frequency comb in arm locking has not been reported.

Here, we describe an updated version of single arm locking, where the noise amplification at the nulls in the science band can be well circumvented by careful control of the noise of the signals. Optical frequency comb is used to precisely divide the laser phase noise by a certain factor into the microwave domain, and the obtained signals, which are immune with the comb frequency noises, are electrically mixed to further generate the arm-locking sensor. We demonstrate the principle of this new kind of single arm locking, and examine the performance of the noise suppression. The results show that, the requirement of TDI can be well satisfied.

\section{Updated single arm locking with precise control of the laser phase noise}\label{section2}
\begin{figure}[htbp]
\includegraphics[width=0.50\textwidth]{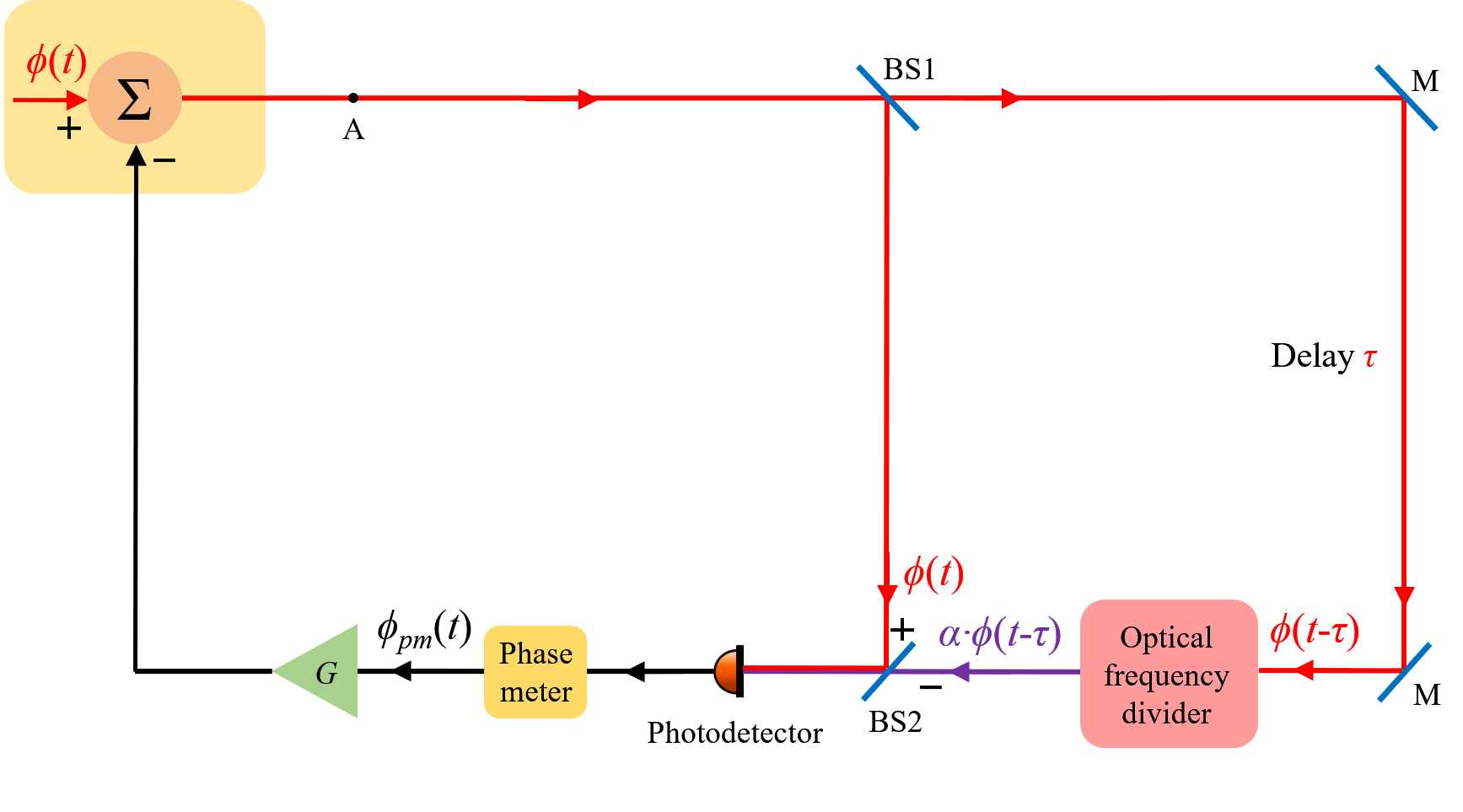}
\caption{\label{fig1}Schematic of the updated single arm locking.}
\end{figure}
At first, we briefly introduce the principle of the updated single arm locking. As shown in Fig. 1, the laser output, whose phase noise is initially ${\phi ({t})}$, is split into two parts at BS1. One part is the local oscillator. The other part is delayed with the time of flight ${\tau}$ (i.e., the incoming beam), and noted as ${\phi (t - \tau )}$. In our scheme and particularly, an optical frequency divider is involved into the incoming beam, which actually serves as an optical frequency synthesizer. In this case, the laser phase noise can be scaled down by a certain factor, and therefore the laser phase noise of the output of the optical frequency divider can be expressed as ${\alpha \cdot \phi (t - \tau)}$, with the downcoversion factor ${\alpha < 1}$. The two beams are combined at BS2, and detected by a photodetector linked to a phase meter. The output signal ${\phi_{pm}(t)}$ of the phase meter is injected into the arm-locking controller, which finally feedback adjusts the laser frequency to reduce the phase noise.

The output signal ${\phi _{pm}}$ of the phase meter can be written as
\begin{equation}\label{N1}
\phi _{pm}(t) = \phi(t) - \alpha \cdot \phi(t - \tau ),
\end{equation}

In the frequency domain, Eq. (1) can be transformed into:
\begin{equation}\label{N2}
{\phi _{pm}}\left( \omega  \right) = \phi \left( \omega  \right) - \alpha  \cdot \phi \left( \omega  \right) \cdot {e^{ - i\omega \tau }},
\end{equation}

When the feedback loop is closed, the phase noise at point A can be expressed as:
\begin{equation}\label{N3}
{\phi _A}\left( \omega  \right) = \frac{{\phi \left( \omega  \right)}}{{1 + G\left( \omega  \right) \cdot \left( {1 - \alpha  \cdot {e^{ - i\omega \tau }}} \right)}}.
\end{equation}
where ${G(\omega)}$ and ${P(\omega) = 1- \alpha  \cdot e^{-i \omega t}}$ are the frequency responses of the controller and the arm locking sensor, respectively. If without the optical frequency divider, the traditional response of the single arm-locking sensor is ${1- e^{-i \omega t}}$. There are a series of nulls at frequencies of ${N / \tau}$, leading to a reduction of the noise suppression in the science band. ${N}$ is an integer. In this work, the updated response goes to ${1- \alpha  \cdot e^{-i \omega t}}$, with ${\alpha < 1}$, whose modulus can not be equal to zero any more, because ${|e^{-i \omega t}|}$ can never be larger than 1. This implies that, the nulls originally located at ${N / \tau}$ have been removed across the whole science band in the configuration of single arm locking. It is obvious that, the ${\alpha}$ value will affect the frequency response of the single arm-locking sensor. Figure 2 shows the Nyquist and Bode diagram for different ${\alpha}$, which are 1, 0.8, 0.5, and 0.1, respectively, in the band from 0.1 mHz to 10 Hz.

We find that, the Nyquist diagram is circular, with (1,0) as the center. When decreasing the value of ${\alpha}$, the radius of the Nyquist diagram decreases. It is clear that, the real part of the sensor can not be equal to zero, when ${\alpha}$ is less than 1. For all the ${\alpha}$ values, the systems are unconditionally stable. Let us focus the Bode diagram. It is well known that, when ${\alpha}$ equals to 1, the magnitude rolls up from zero to 2, and then rolls off. The phase features the differentiator (advanced phase) and the integrator (lag phase), sequentially, and experiences a jump from ${-\pi /2}$ to ${\pi /2}$ at each nulls. In sharp contrast, with ${\alpha < 1}$, the magnitude can not reach zero as the lower bound, which changes in the range from ${1 - \alpha}$ to ${1 + \alpha}$. This implies that, the noise amplification at the nulls can not reach infinite, and the peaks could be inhibited. In the case of the phase frequency characteristics, the phase becomes more stable when ${\alpha}$ is less than 1. When ${\alpha}$ is set to 0.1, the phase change range is only from ${-\pi /20}$ to ${\pi /20}$. We consider that, this fact could relax the phase compensation following in the design of the arm-locking controller.

\begin{figure*}[htbp]
\centering
\subfigure{%
    \includegraphics[width=3in]{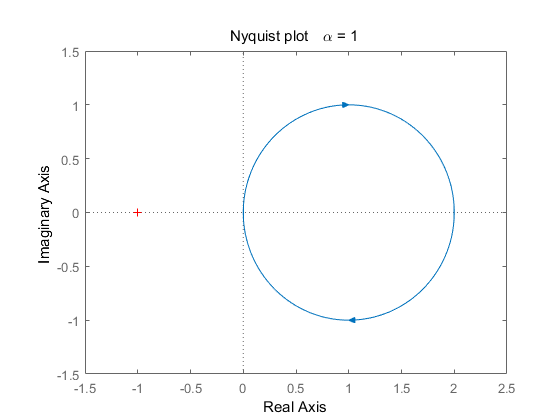}}
\quad
\subfigure{
    \includegraphics[width=3in]{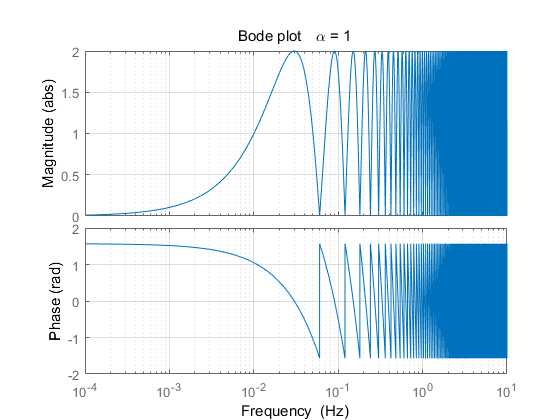}}

\subfigure{%
    \includegraphics[width=3in]{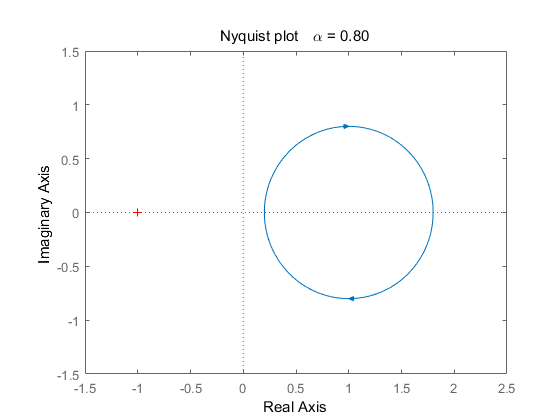}}
\quad
\subfigure{
    \includegraphics[width=3in]{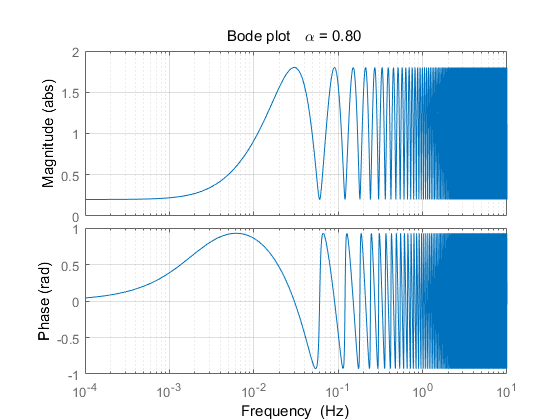}}

\subfigure{%
    \includegraphics[width=3in]{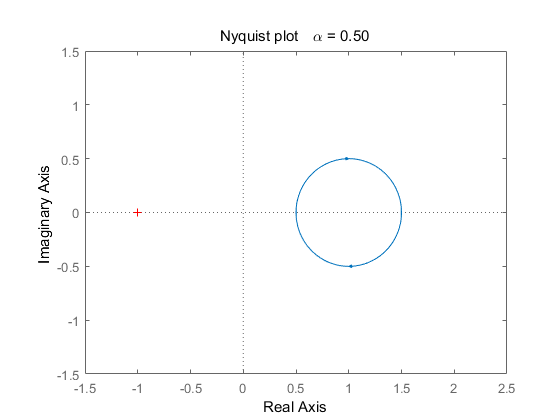}}
\quad
\subfigure{
    \includegraphics[width=3in]{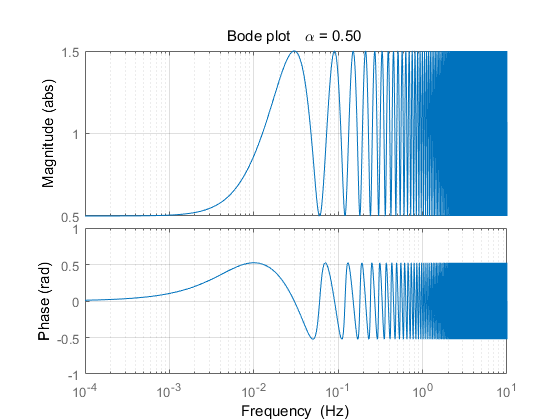}}

\subfigure{%
    \includegraphics[width=3in]{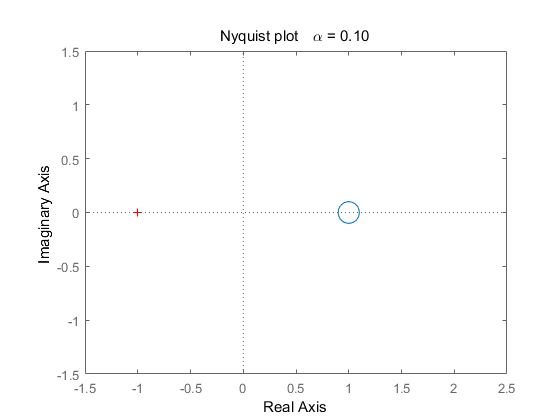}}
\quad
\subfigure{
    \includegraphics[width=3in]{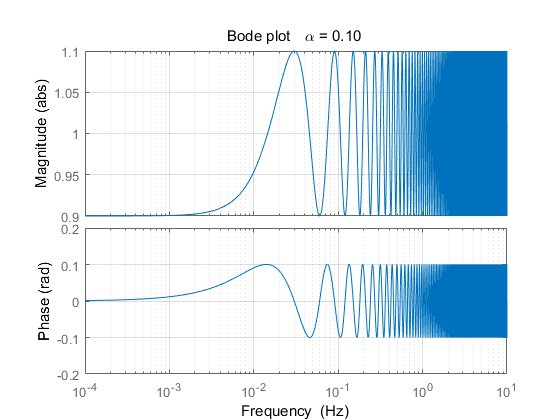}}
\caption{\label{fig2} Nyquist diagram and Bode diagram of the single arm-locking sensor with different ${\alpha}$ values.}
\end{figure*}

\section{Updated single arm locking with optical frequency comb}\label{section3}
\subsection{$\textbf{Precise control of the laser phase noise}$}
In Sec. 2, we use the optical frequency divider to scale down the laser phase noise for convenience of the explanation of the updated single arm locking. However, limited by the state of the art of the photodetector bandwidth, ${\alpha}$ can not be sufficiently small. For example, the 300 GHz bandwidth corresponds to the ${\alpha}$ value of about 0.9989, when the laser wavelength is 1064 nm. This means, if the frequencies of the local oscillator and the incoming light are far different from each other, we can not observe the beat note at the photodetector. It is obvious that, smaller ${\alpha}$ value can result in a better suppression of the peak (Please note that, ${\alpha}$ can not be zero, since if that there is not returned signal from the distant spacecraft. We will discuss the bound of the ${\alpha}$ value hereafter). Therefore, one challenge is emerging, which is how to coherently make ${\alpha}$ small enough. One solution could be the technique of optical frequency comb.

Optical frequency combs are able to work as a bidirectional bridge, with the optical frequency and the microwave frequency standing at the two sides. We can use frequency comb to coherently transfer the optical noise into the microwave regime, with an appropriate downconversion factor [29]. Consequently, the mixing of the microwave signals could become easier, which actually corresponds to the heterodyne beat of the optical signals. By scaling up, the electrical mixing signal can serve as the input of the arm-locking controller to suppress the phase noise of the laser source, after the possible transfer to the phase term by an integrator.

\begin{figure*}[htbp]
\includegraphics[width=0.60\textwidth]{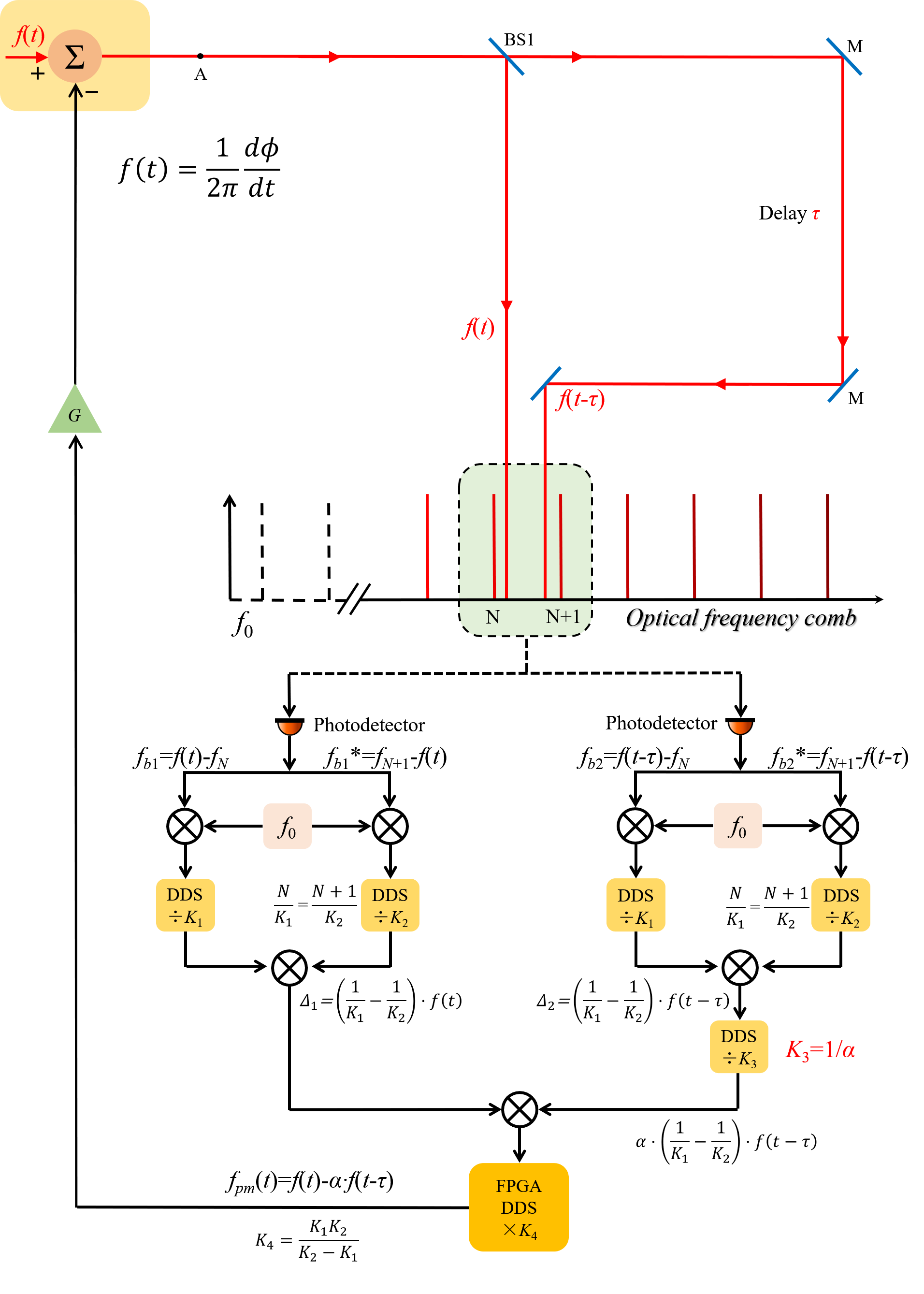}
\caption{\label{fig3} Schematic of the laser phase noise control using optical frequency comb.}
\end{figure*}

For convenience of explanation, we do not involve the clock noise, the spacecraft motion, and the shot noise into the derivation, and the distant spacecraft just works as a perfect mirror in this subsection. The detailed method is shown in Fig. 3. The local oscillator ${f(t)}$ beats with the two nearby comb lines (e.g., the ${N}$th and ${N+1}$th line), and the beat notes can be expressed as:
\begin{equation}\label{N4}
{f_{b1}} = f\left( t \right) - {f_N} = f\left( t \right) - N \cdot {f_{rep}} - {f_0},
\end{equation}
\begin{equation}\label{N5}
{f_{b1}}^* = {f_{N + 1}} - f\left( t \right) = \left( {N + 1} \right) \cdot {f_{rep}} + {f_0} - f\left( t \right),
\end{equation}
where ${f_N}$ is ${N \times f_{rep} + f_0}$, ${f_{N+1}}$ is ${(N+1) \times f_{rep} + f_0}$, ${N}$ is an integer, ${f_{rep}}$ is the repetition frequency, and ${f_0}$ is the carrier envelope offset frequency of the frequency comb. To start with, the two beat frequencies are mixed with ${f_0}$, and can be updated to:
\begin{equation}\label{N6}
{f_{b1}}^\prime  = f\left( t \right) - N \cdot {f_{rep}},
\end{equation}
\begin{equation}\label{N7}
{f_{b1}^{* \prime }} = \left( {N + 1} \right) \cdot {f_{rep}} - f\left( t \right).
\end{equation}

The mode number ${N}$ and ${N+1}$ can be precisely measured with uncertainty better than 0.5, by a wavelength meter. ${{f_{b1}}}$ and ${{f_{b1}^{* \prime }}}$ are then divided by ${K_1}$ and ${K_2}$ based on two direct digital synthesizers (DDS), where
\begin{equation}\label{N8}
\frac{N}{{{K_1}}} = \frac{{N + 1}}{{{K_2}}},
\end{equation}

The outputs of the two DDS are then mixed, and we can get
\begin{equation}\label{N9}
{\Delta _1} = \left( {\frac{1}{{{K_1}}} - \frac{1}{{{K_2}}}} \right) \cdot f\left( t \right),
\end{equation}

We find that, the signal ${\Delta _1}$ in the electrical domain is immunity to the comb frequency noise, and can be coherently traceable to the local oscillator ${f(t)}$ with a downconversion factor of ${(K_2-K_1)/K_1 K_2}$.

In the similar way, we can get another signal corresponding to the incoming beam from the distant spacecraft, as:
\begin{equation}\label{N10}
{\Delta _2} = \left( {\frac{1}{{{K_1}}} - \frac{1}{{{K_2}}}} \right) \cdot f\left( {t - \tau } \right),
\end{equation}

Then, an additional DDS is added into the signal link, which is of key importance in the configuration of the updated single arm locking. ${\Delta _2}$ is divided by ${K_3}$, ${K_3=1/ \alpha}$, and we achieve
\begin{equation}\label{N11}
\begin{aligned}
{\Delta _3} &= \frac{1}{{{K_3}}} \cdot \left( {\frac{1}{{{K_1}}} - \frac{1}{{{K_2}}}} \right) \cdot f\left( {t - \tau } \right) \\
&= \alpha  \cdot \left( {\frac{1}{{{K_1}}} - \frac{1}{{{K_2}}}} \right) \cdot f\left( {t - \tau } \right),
\end{aligned}
\end{equation}
\begin{figure*}[htbp]
\includegraphics[width=0.90\textwidth]{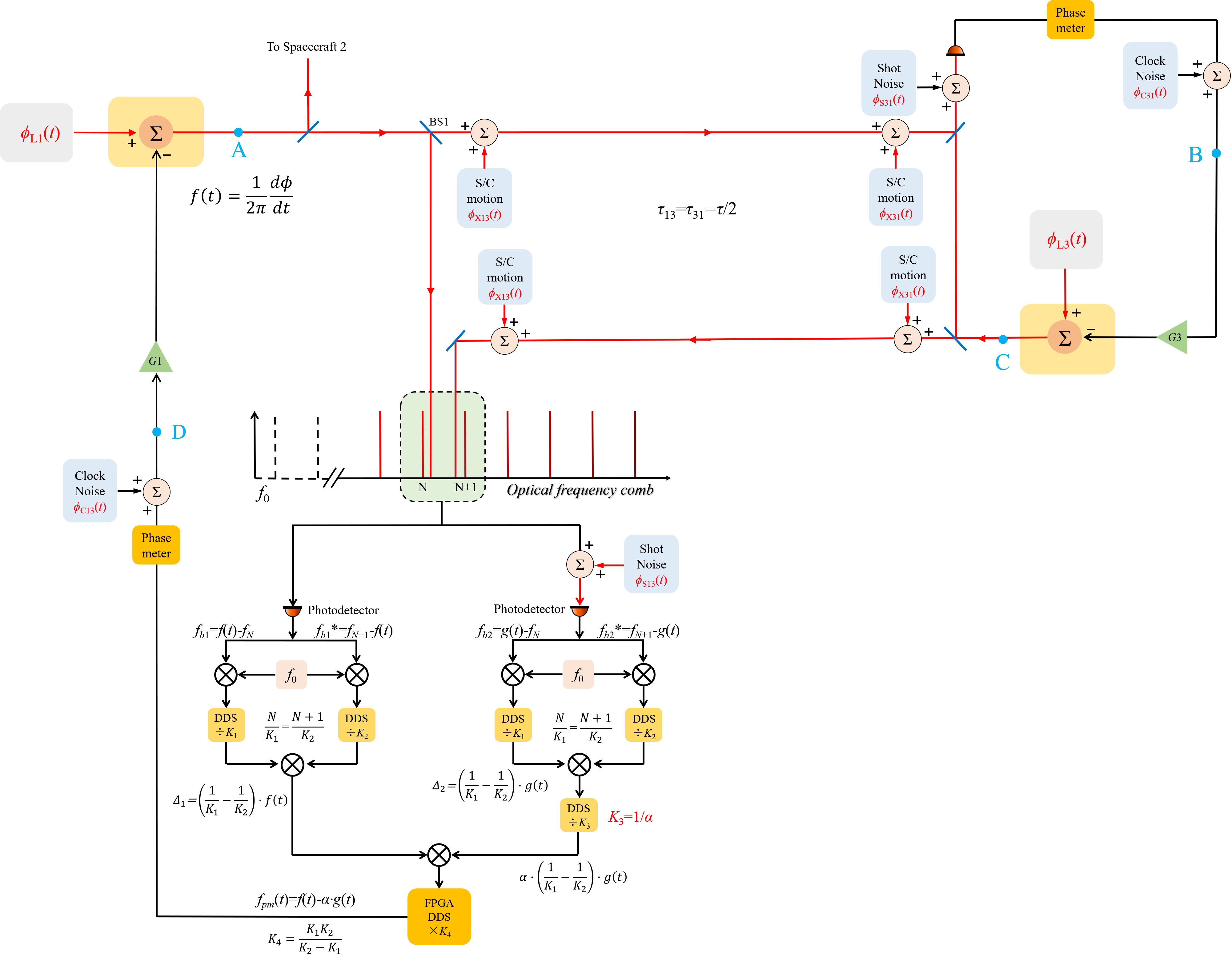}
\caption{\label{fig4} Schematic of the single arm locking with optical frequency comb.}
\end{figure*}

Based on Eq. (11), we find that, the laser phase noise can be coherently and arbitrarily scaled with a certain factor by the virtue of the optical-to-electrical network. This is rather difficult in the optical domain actually. ${\Delta _1}$ and ${\Delta _3}$ are mixed, and then multiplied by ${K_4}$. ${K_4=K_1K_2/(K_2-K_1)}$. Please note that, ${K_4}$ could be very large, and the commercial devices could not meet the requirement. However, this DDS can be customized using FPGA. The final signal can be expressed as:
\begin{equation}\label{N12}
{f_{pm}}\left( t \right) = f\left( t \right) - \alpha  \cdot f\left( {t - \tau } \right),
\end{equation}
In the phase term (i.e., both sides multiplied by ${2\pi t}$), Eq. (12) can be rewritten as:
\begin{equation}\label{N13}
{\phi _{pm}}\left( t \right) = \phi \left( t \right) - \alpha  \cdot \phi \left( {t - \tau } \right).
\end{equation}
which can be sent into the arm-locking controller to feedback control the laser frequency.

The parameters can be easily determined. We assume that the laser wavelength is 1064 nm in vacuum, i.e., about 281.759 THz. Recently, the dissipative Kerr soliton (DKS) combs have found a wealth of applications. If the repetition frequency and the carrier-envelope-offset frequency are 5 GHz and 2 GHz, respectively. ${f_{b1}}$ can be measured to be 2.8 GHz, and ${{f_{b1}^{*}}}$ is therefore 2.2 GHz. ${N}$ can be determined as 56351. If we set ${K_1}$ as 200, ${K_2}$ can be calculated as 200.0035. ${\Delta _1}$ is then roughly 27.9 MHz, while ${\Delta _2}$ is at the same level. In this case, ${K_4}$ is about ${1.1 \times 10^{7}}$, which can be realized via a FPGA-based signal processor.
\begin{figure*}[htbp]
\centering
\subfigure{%
    \includegraphics[width=3in]{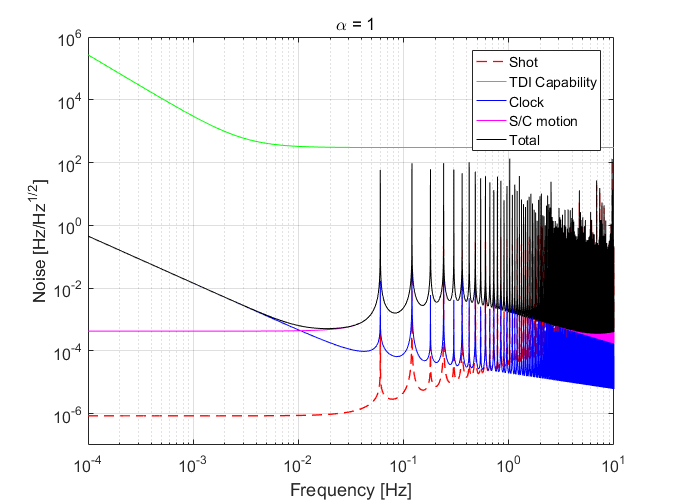}}
\quad
\subfigure{%
    \includegraphics[width=3in]{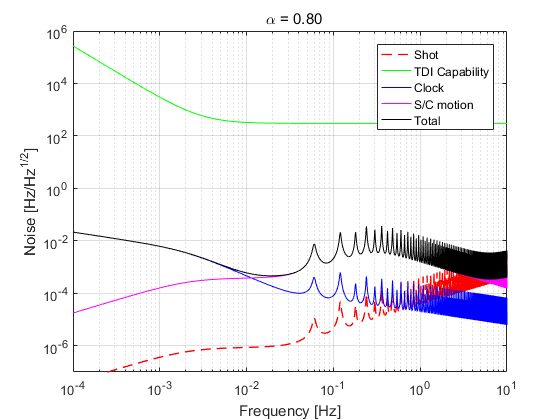}}

\subfigure{
    \includegraphics[width=3in]{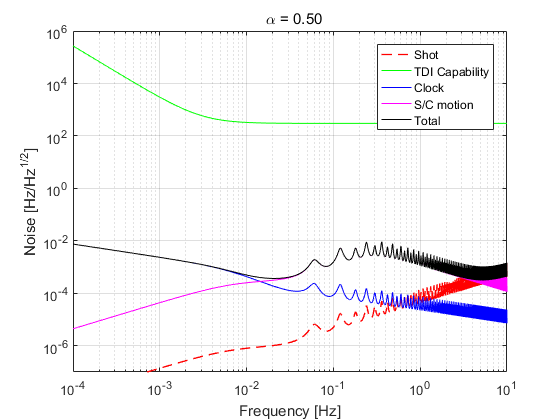}}
\quad
\subfigure{%
    \includegraphics[width=3in]{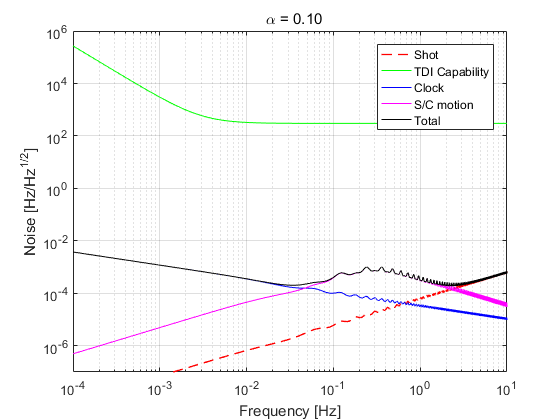}}

\caption{\label{fig5} Noise floor of updated single arm-locking at the laser output with different ${\alpha}$ values. The total noise comprises clock noise, spacecraft motion, and shot noise.}
\end{figure*}

\subsection{$\textbf{Updated single arm locking with optical frequency comb}$}
In this subsection, we introduce the clock noise, the spacecraft motion, the shot noise, and the distant spacecraft into the system. The schematic is shown in Fig. 4. Along the line in Ref. [11], the clock noise, the noise due to the spacecraft motion, and the shot noise can be estimated as:
\begin{equation}\label{N14}
{\phi _{Cij}}\left( f \right) = {f_{Dij}}{C_i}\left( f \right) = {f_D}\frac{{{y_i}\left( f \right)}}{{2\pi f}},
\end{equation}
\begin{equation}\label{N15}
{\phi _{Xij}}\left( f \right) = \frac{{\Delta {X_{ij}}\left( f \right)}}{\lambda },
\end{equation}
\begin{equation}\label{N16}
{\phi _{Sij}} = {\left( {\frac{{\hbar c}}{{2\pi }}\frac{1}{{\lambda {P_d}}}} \right)^{{1 \mathord{\left/
 {\vphantom {1 2}} \right.
 \kern-\nulldelimiterspace} 2}}}.
\end{equation}
where ${\phi _{Cij}}$ is the clock noise at the ${i}$th spacecraft, associated with the incoming light from the ${j}$th spacecraft, ${f_{Dij}}$ is the Doppler frequency, ${{C_{i}(f)}}$ is the normalized clock noise, and ${y_{i}(f)}$ is the fractional frequency fluctuations of the clock. ${\phi _{Xij}}$ is the phase noise caused by the spacecraft motion, ${\Delta {X_{ij}}}$ is the motion displacement facing to the ${j}$th spacecraft at the ${i}$th spacecraft, and ${\lambda}$ is the laser wavelength. ${\phi _{Sij}}$ is the shot noise due to the quantization nature of light, ${\hbar}$ is the Planck constant, ${c}$ is the light speed in vacuum, and ${P_d}$ is the light power from the distant spacecraft. The unit for each noise is ${\rm cycles /Hz^{1/2}}$ in Eqs. (14), (15), and (16). Please note that, the shot noise in the local branch (i.e., the branch of ${f(t)}$) can be neglected, since the power of the local oscillator is sufficiently high.

The phase noise at the phase meter output on spacecraft 3 is given by:
\begin{equation}\label{N17}
\begin{aligned}
\phi _B(\omega) &= {\phi _{L3}}(\omega) - {\phi _{L3}}(\omega){e^{ - i\omega {\tau _{13}}}} + {\phi _{X13}}( \omega){e^{ - i\omega {\tau _{13}}}} \\
 & + {\phi _{X31}}(\omega) + {\phi _{S31}}( \omega ) + {\phi _{C31}}( \omega ),
\end{aligned}
\end{equation}

When the phase locking loop is closed on spacecraft 3, the phase noise at point C is
\begin{equation}\label{N18}
\begin{aligned}
\phi _C ( \omega) &= \frac{{\phi _{L3}(\omega)}}{{1 + {G_3}(\omega)}} + \frac{{{G_3}(\omega)}}{{1 + {G_3}( \omega )}} \left (
{\phi _{L1}}( \omega ){e^{ - i\omega {\tau _{13}}}} \right.\\
 &+ {\phi _{X13}}( \omega ){e^{ - i\omega {\tau _{13}}}}+ {\phi _{X31}}( \omega )\\
 &- \left. {\phi _{S31}}( \omega) - {\phi _{C31}}(\omega) \right),
\end{aligned}
\end{equation}

Please note that, the gain ${G_3}$ is much larger than 1. Therefore, the coefficients of ${1/(1+G_3)}$ and ${G_3/(1+G_3)}$ are approximately equal to 0 and 1, respectively. The phase noise at the phase meter output on the spacecraft 1 is given by
\begin{equation}\label{N19}
\begin{aligned}
\phi _D(\omega )
 & = {\phi _{L1}}( \omega )( {1 - \alpha {e^{ - i\omega \tau }}} ) - \alpha  \cdot [ {\phi _{X13}}( \omega  )( {1 + {e^{ - i\omega \tau }}} ) \\
 & + 2{\phi _{X31}}( \omega  ){e^{ - i\omega {\tau _{31}}}}- {\phi _{S31}}( \omega  ){e^{ - i\omega {\tau _{31}}}} \\
 & - {\phi _{C31}}( \omega  ){e^{ - i\omega {\tau _{31}}}}]+ { \alpha } \cdot {\phi _{S13}}( \omega ) + {\phi _{C13}}(\omega),
\end{aligned}
\end{equation}

We find that, the first term is related to the laser phase noise ${\phi _{L1}}$, and the transfer function has been updated to ${1- \alpha  \cdot e^{-i \omega t}}$. All the other noises in the incoming beam has been scaled by a factor of ${\alpha}$. When the loop on the spacecraft 1 is closed, the phase noise at point A can be calculated as
\begin{equation}\label{N20}
\begin{aligned}
\phi _{A,CL}(\omega) &= \frac{{{\phi _{L1}}( \omega)}}{{1 + {G_1}( \omega){P(\omega)}}} - \frac{{{G_1}(\omega)}}{{1 + {G_1}(\omega)( {P(\omega)})}}[ - \alpha  \cdot \\
  & [{\phi _{X13}}( \omega )( {1 + {e^{ - i\omega \tau }}}) + 2{\phi _{X31}}( \omega){e^{ - i\omega {\tau _{31}}}} \\
  & - {\phi _{S31}}( \omega  ){e^{ - i\omega {\tau _{31}}}} - {\phi _{C31}}(\omega){e^{ - i\omega {\tau _{31}}}}]\\
  & + \alpha \cdot {\phi _{S13}}(\omega) + {\phi _{C13}}(\omega)].
\end{aligned}
\end{equation}

Armed with Eq. (20), we can obtain the noise floor of updated single arm locking. Firstly, we neglect laser phase noise, namely, the controller has infinite gain. The design of controller and the total noise budget, including laser noise, will be discussed in Sec. 4. In the case of high controller gain, noise sources other than laser phase noise have little dependence on the controller gain, depending primarily on the function of the arm-locking sensor. Assuming the arm length is ${2.5 \times 10^9}$ m (8.3 s in time), the noise floors of updated single arm locking are plotted in Fig. 5 for different ${\alpha}$, respectively, in the band from 0.1 mHz to 10 Hz.

Also plotted in Fig. 5 is the expected TDI capability (green curve), equal to the frequency noise requirement before TDI, which is given by
\begin{equation}\label{N21}
{v_{{\rm{TDI}}}}(f) \approx 300 \times ( {1 + {{\left( {\frac{{3{\kern 1pt} {\kern 1pt} {\rm{mHz}}}}{f}} \right)}^2}}) {\kern 1pt} {\kern 1pt} {\kern 1pt} {\kern 1pt} \frac{{{\rm{Hz}}}}{{\sqrt {{\rm{Hz}}} }}.
\end{equation}

We find that, when ${\alpha}$ equals to 1, the arm-locking sensor works as the traditional single arm locking. There are a series of the peaks at the nulls, which actually corresponds to the infinite noise amplification. In addition, the noise floor is relatively steep during the whole science band (${0.45 \, \rm Hz/Hz^{1/2}}$ @ 0.1mHz, ${6.42 \, \rm Hz/Hz^{1/2}}$ @ 10mHz, ${1.8 \, \times 10^{-3} \rm Hz/Hz^{1/2}}$ @ 1Hz ). With decreasing the ${\alpha}$ value, the peaks are becoming smooth, whose magnitudes are proportional to the coefficient of ${1/(1-\alpha)}$. For example, the ${\alpha}$  value of 0.5 means 2 times of amplification of the noises at the nulls. When ${\alpha}$ is reduced to 0.95, the peaks can not reach the curve of the TDI capability already. As ${\alpha}$ is as small as 0.1, the noise floor becomes relatively flat (${3.6 \times 10^{-3} \, \rm Hz/Hz^{1/2}@0.1mHz}$, ${3.5 \times 10^{-4} \, \rm Hz/Hz^{1/2}@10mHz}$, ${4 \times 10^{-4} \, \rm Hz/Hz^{1/2}@1Hz}$), and all below ${10^{-2}\rm \, Hz/Hz^{1/2}}$ across the science band. This is because the clock noise and the noise due to the spacecraft motion from the distant spacecraft has been scaled down by ${\alpha}$, as shown in Eq. (20). It is clear that a smaller ${\alpha}$ is preferable. We can evaluate the upper bound of ${\alpha}$ via the noise magnitude at the nulls (i.e., the peaks), whose band corresponds to the noise caused by the spacecraft motion. If we assume that the noise floor should be below ${1 \rm \, Hz/Hz^{1/2}}$, this condition is given by
\begin{equation}\label{N22}
\frac{{{\phi _{Xij}}\left( f \right) \cdot 2\pi f}}{{1 - \alpha }} < 1{{{\kern 1pt} {\kern 1pt} {\kern 1pt} {\rm{Hz}}} \mathord{\left/
 {\vphantom {{{\kern 1pt} {\kern 1pt} {\kern 1pt} {\rm{Hz}}} {\sqrt {{\rm{Hz}}} }}} \right.
 \kern-\nulldelimiterspace} {\sqrt {{\rm{Hz}}} }},
\end{equation}

Therefore, the maximum value of ${\alpha}$ can be estimated as:
\begin{equation}\label{N23}
\alpha  < 1 - {\phi _{Xij}}\left( f \right) \cdot 2\pi f,
\end{equation}

Consider the lower bound of ${\alpha}$, it is obvious that ${\alpha}$ can not be infinitely small, since there is not the delayed length information in that case. Based on Fig. 4, the design of ${\alpha}$ should maintain that the frequency ${\alpha  \cdot \left( {\frac{1}{{{K_1}}} - \frac{1}{{{K_2}}}} \right) \cdot g\left( t \right)}$ can be distinguished by the following electrical network, while it is reasonable that the frequency resolution can achieve 1 mHz for most of the electrical devices. ${g(t)}$ is the optical frequency of the incoming beam. Therefore, we reach
\begin{equation}\label{N24}
\alpha  \cdot \left( {\frac{1}{{{K_1}}} - \frac{1}{{{K_2}}}} \right) \cdot g\left( t \right) > 1{\kern 1pt} {\kern 1pt} {\rm{mHz}},
\end{equation}

The minimum value of ${\alpha}$ can be thus estimated as:
\begin{equation}\label{N25}
\alpha  > \frac{{1{\kern 1pt} {\kern 1pt} {\rm{mHz}}}}{{\left( {\frac{1}{{{K_1}}} - \frac{1}{{{K_2}}}} \right) \cdot g\left( t \right)}}.
\end{equation}

Along this guideline, the ${\alpha}$ value, greater than 0.1 and less than 0.8, is suggested.
\begin{figure*}[htbp]
\centering
\subfigure[]{%
    \includegraphics[width=3.5in]{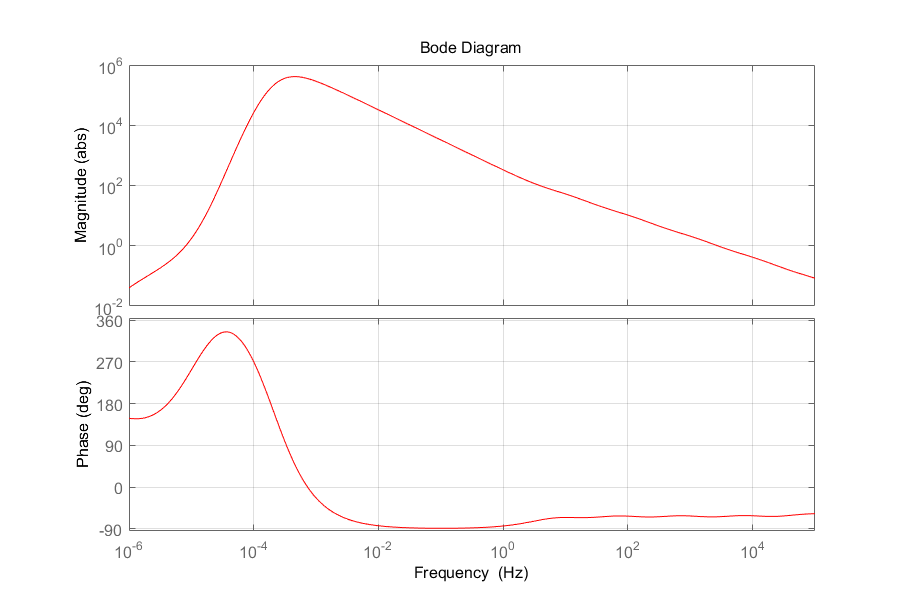}}
\quad
\subfigure[]{
    \includegraphics[width=3in]{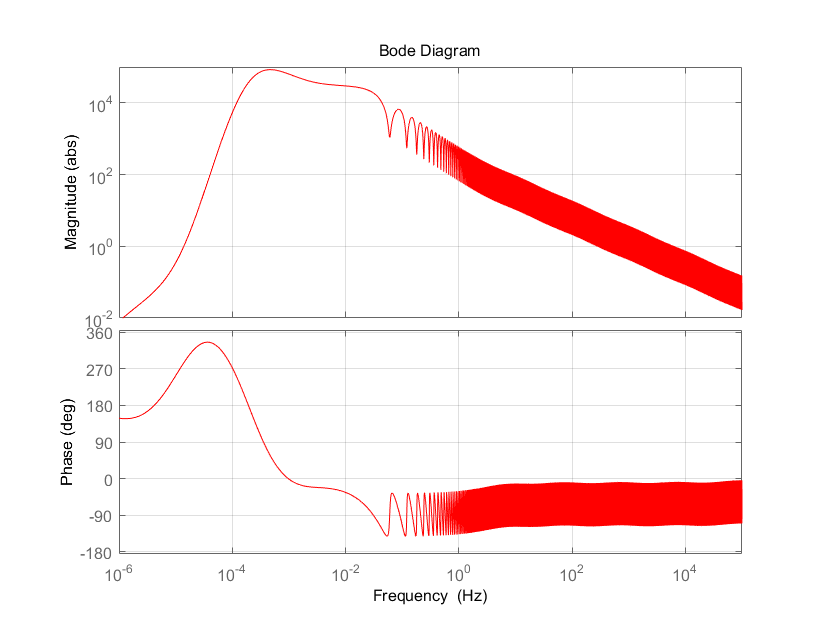}}

\caption{\label{fig6}  a) Bode plot of arm-locking controller ${G \cdot {\phi _{pm}}\left( t \right)}$ and b) open-loop gain   with ${\alpha}$ = 0.80.}
\end{figure*}
	
\section{ Performance of updated single arm locking with optical frequency comb}
\subsection{Design of the updated single arm locking controller}
The arm locking controller is required to provide sufficient gain to suppress the laser phase noise to be below the TDI capability curve within the science band. The design of controller can be divided into two parts: low-frequency part (< 1 mHz) and high-frequency part (> 1 Hz). For the low-frequency part, to control the Doppler frequency pulling at low-frequency, the controller is required to be AC coupled with high-pass filtering at very low frequencies (~10 ${\rm \mu Hz}$). The high-pass filter is based on a classical lead-lag design. Similar with the discussion in Ref. [11], the cut-off frequency is design to be ${2.2 \times 10^{-4} \rm Hz}$ and a unity-gain frequency around 10 ${\rm \mu Hz}$ is introduced (0 dB@ 8.2${\, \rm \mu Hz}$ and 107 dB@${\, 2.3 \times 10^{-4}}$Hz). For the design at frequencies in the science band, we implement a low pass filter at ${1 \times 10^{-4}}$ Hz to limit the controller action. In addition, practical bandwidth limitations require that the loop gain roll-off at several kHz. Therefore, a slope of magnitude about ${f^{-2/3}}$ is obtained to achieve a high gain in the science band (112 dB@${5.6 \times 10^{-4}}$Hz and 0 dB@2.4kHz). This can be achieved by placing zeros and poles alternately with a frequency spacing ratio of 10, i.e., poles at 10 Hz, 100 Hz, 1 kHz, 10 kHz, 100 kHz and zeros at 5 Hz, 50 Hz, 500 Hz, 5 kHz, 50 kHz, 500 kHz, which is also sufficient to maintain enough phase margin. Figure 6 a) contains the Bode plot of the controller. Now looking at the controller phase, it changes smoothly in the science band and high frequencies, which provides a better stability for the closed-loop system. The gain margin and phase margin of our controller are 3.64 and ${52.7 ^{\circ}}$, respectively. Furthermore, we also plot the open-loop gain   with ${\alpha}$ = 0.80 in Figure 6 b). The unity-gain frequencies at low-frequency and high-frequency are 8.2 ${\rm \mu Hz}$ and 2.4 kHz, respectively. For other ${\alpha}$ values, the results are similar, which can limit the Doppler frequency pulling and meet the requirement of bandwidth limitations.

\subsection{Performance assuming free-running laser noise and pre-stabilized laser noise}
In this section, we use three types of laser frequency noises, which are free-running (FR) laser noise, Mach-Zehnder (MZ) stabilization, and Fabry-Perot (FP) cavity stabilization [11]. The laser frequency noises can be expressed as:

\begin{equation}\label{N26}
{\kern 1pt} {\nu _{FR}}\left( f \right) = 30000 \times \frac{{1{\kern 1pt} {\kern 1pt} {\rm{Hz}}}}{f}{\kern 1pt} {\kern 1pt} {\kern 1pt} {\kern 1pt} {\kern 1pt} {\rm{Hz}}/\sqrt {{\rm{Hz}}}.
\end{equation}
\begin{equation}\label{N27}
\begin{array}{*{20}{c}}
{{\nu _{MZ}}\left( f \right) = 800 \times \left( {1 + {{\left( {\frac{{2.8{\kern 1pt} {\kern 1pt} {\rm{mHz}}}}{f}} \right)}^2}} \right)}&{{\rm{Hz}}/\sqrt {{\rm{Hz}}} }
\end{array}.
\end{equation}
\begin{equation}\label{N28}
\begin{array}{*{20}{c}}
{{\nu _{FP}}\left( f \right) = 30 \times \left( {1 + {{\left( {\frac{{2.8{\kern 1pt} {\kern 1pt} {\kern 1pt} {\rm{mHz}}}}{f}} \right)}^2}} \right)}&{{\rm{Hz}}/\sqrt {{\rm{Hz}}} }
\end{array}.
\end{equation}

Figure 7 shows the noise budget of updated single arm-locking with free-running laser noise. Benefitting from the introduce of the coefficient ${\alpha}$, the peaks in the science band is not infinite. The peaks get slighter with smaller ${\alpha}$. We can find that the laser frequency noise is the limiting noise source (the system is gain limited) with the other system noise sources well below the laser frequency noise. It also shows that even without any form of laser pre-stabilization, arm locking will meet the TDI capability across the entire science band while ${\alpha}$ < 1. The noise level is above the floor shown in Fig. 5, due to the limited gain in practice. Compared with the modified dual arm-locking sensor, this updated single arm-locking sensor is able to work continuously for the full year, and the noise suppression performance is not related to the arm length mismatch, showing a high robustness.

\begin{figure*}[htbp]
\centering
\subfigure{%
    \includegraphics[width=2.5in]{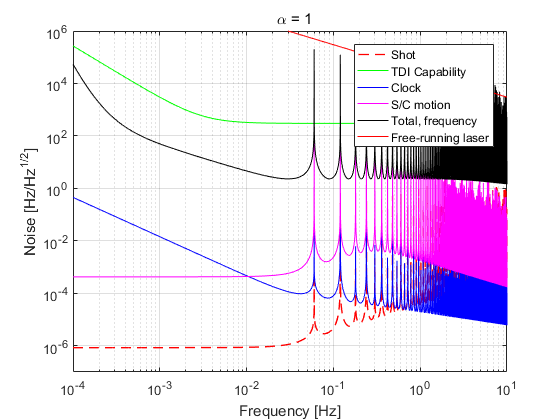}}
\quad
\subfigure{%
    \includegraphics[width=2.5in]{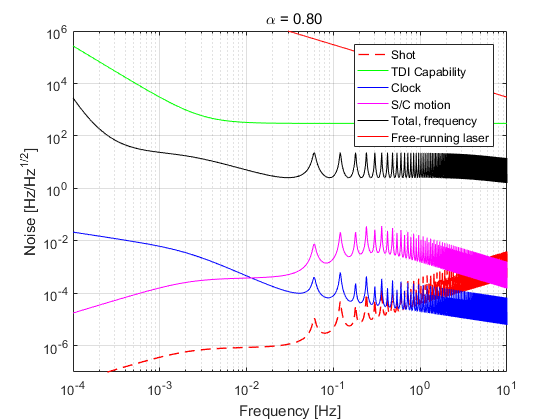}}

\subfigure{
    \includegraphics[width=2.5in]{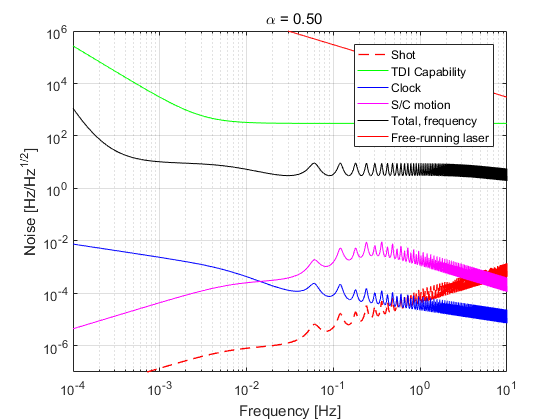}}
\quad
\subfigure{%
    \includegraphics[width=2.5in]{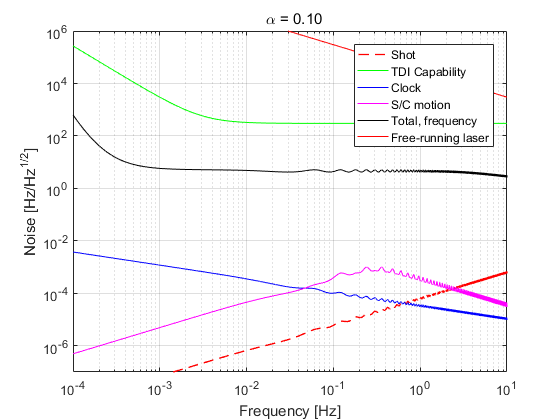}}

\caption{\label{fig7} The noise budget of updated single arm-locking with different ${\alpha}$ values. The performance was calculated with free-running laser noise as an initial condition.}
\end{figure*}
\begin{figure*}[htbp]
\centering
\subfigure{%
    \includegraphics[width=2.5in]{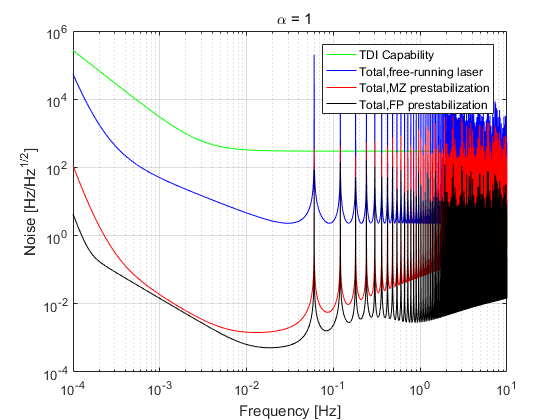}}
\quad
\subfigure{%
    \includegraphics[width=2.5in]{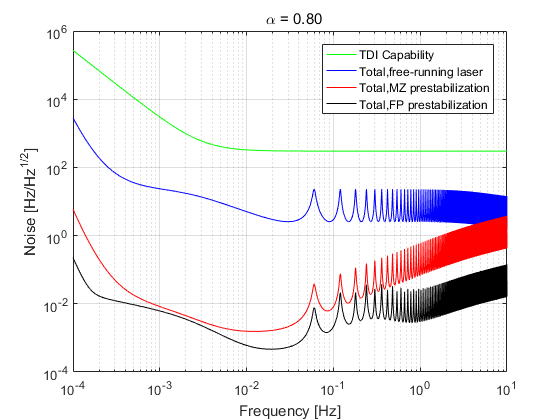}}

\subfigure{
    \includegraphics[width=2.5in]{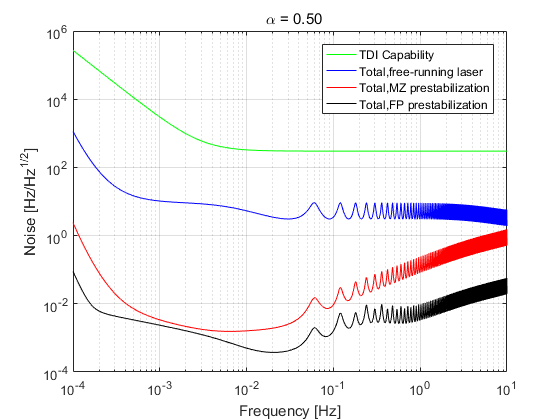}}
\quad
\subfigure{%
    \includegraphics[width=2.5in]{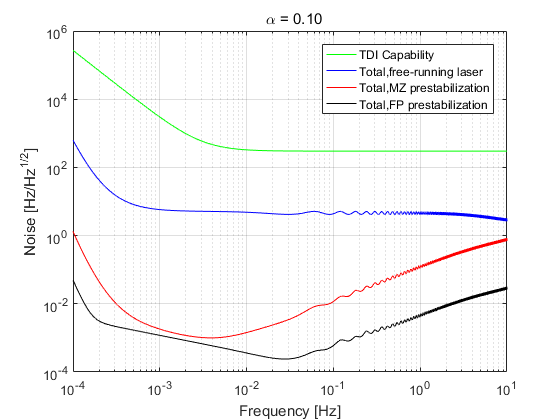}}

\caption{\label{fig8} Noise of updated single arm-locking with different laser frequency noises for different ${\alpha}$ values.}
\end{figure*}

In Fig. 8, the total noise after arm locking is plotted in the cases of no pre-stabilization, Fabry-Perot cavity stabilization, and Mach-Zehnder stabilization. It is clear that either pre-stabilization type in combination with arm locking will deliver performance several orders of magnitude better than the TDI capability.

\begin{figure*}[htbp]
\centering
\subfigure{%
    \includegraphics[width=3in]{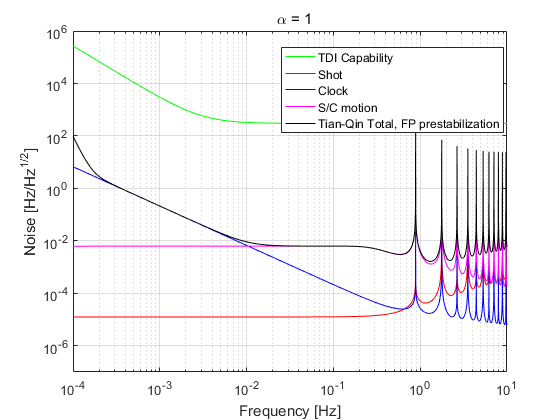}}
\quad
\subfigure{%
    \includegraphics[width=3in]{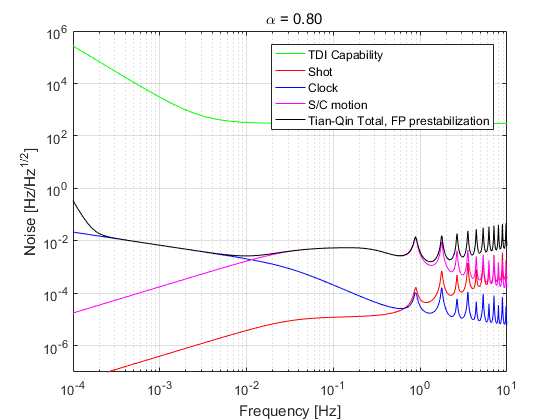}}

\subfigure{
    \includegraphics[width=3in]{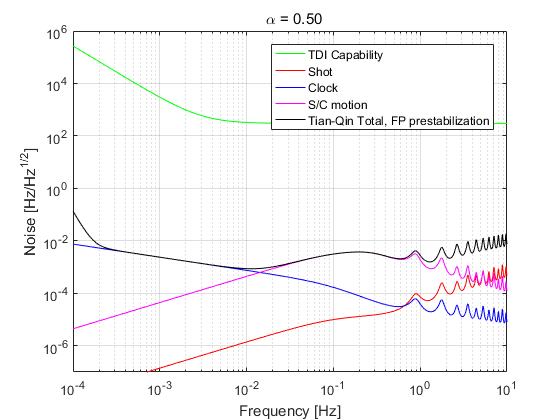}}
\quad
\subfigure{%
    \includegraphics[width=3in]{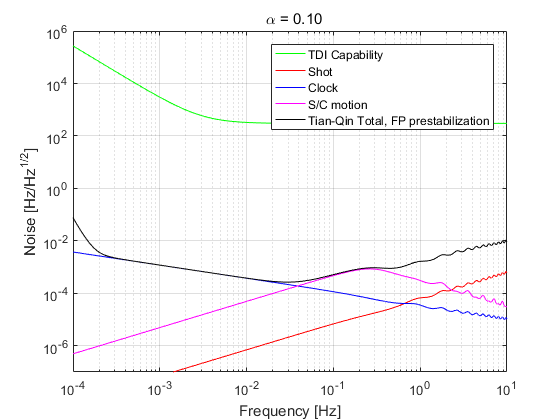}}

\caption{\label{fig9} The noise budget of updated single arm-locking for Tian-Qin with different ${\alpha}$ values.}
\end{figure*}

It is necessary to evaluate the results in the configuration of TianQin, whose baseline is about ${1.7 \times 10^8}$ m (0.57 s in time). The master laser in TianQin will be locked to a Fabry-Perot cavity, whose frequency noise can be expressed as [5]
\begin{equation}\label{N29}
\begin{array}{*{20}{c}}
{{\nu _{FP}^{TQ}}\left( f \right) = 10 \times \left( {1 + {{\left( {\frac{{6{\kern 1pt} {\kern 1pt} {\rm{mHz}}}}{f}} \right)}^2}} \right)}&{{\rm{Hz}}/\sqrt {{\rm{Hz}}} }
\end{array}.
\end{equation}

The results of the noise reduction are shown in Fig. 9. Since the initial condition of the pre-stabilization laser is stable, we find that the total noise can reach the noise floor in a broad band with the current design of the arm-locking controller. In the low frequency below about 0.2 mHz, and high frequency above about 0.5 Hz, the total noise is beyond the floor. This is because the gain of the controller in these bands rolls off gradually, as shown in Fig. 6. In the band from 1 mHz to 0.1 Hz, the residual noise is determined by the clock noise and the spacecraft motion. In fact, we consider that, the clock noise and the spacecraft motion can be reduced by further modifying the arm-locking sensor (discussed in Sec. 5).

\subsection{Frequency pulling estimation}
On the spacecraft, the beat frequencies are measured through the known offset in the phase locking loop at the distant spacecraft and the Doppler frequency caused by the relative spacecraft motion. The phase meter demodulates the beat signal with the difference frequency of the two lasers. In practice, the preset frequency at the phase meter should be updated in real time, since the Doppler frequency is changing all the time. However, the Doppler frequency can not be determined accurately, resulting in a frequency pulling effect to the master laser. Further, all the six lasers will be pulled in the constellation. The solution is to set the arm-locking controller into the AC-coupling mode, which means the gain at low frequencies is small. The frequency pulling in the dual arm locking is significant, because the frequency response to differential Doppler error is inversely proportional to the mismatch of the arm length. Fortunately, the single or common arm locking features a better frequency pulling characteristics. As demonstrated in Ref. [11], at the lock acquisition, the maximum pulling is 460 and 90 MHz, for the free-running and pre-stabilized lasers, respectively. In the case of steady state, the maximum pulling is less than 8 MHz for over 800 days, when the mode is common arm locking. We consider that, the maximum pulling of the updated single arm locking presented here is at the same level as the common arm locking. In general, the mode-hopping free tuning range for NPRO lasers is about 10 GHz, much larger than 460 MHz. Therefore, there is not potential risk of the mode hopping in the operation.

\section{Discussion}\label{section5}
In Sec. 2, only the returned signal from the far spacecraft is tuned by a certain factor, to better understand this presented method. We would like to describe a unified approach in this section, to demonstrate the updated single arm locking comprehensively. Actually, the local oscillator can be also precisely controlled based on the same architecture. The scheme is shown in Fig. 10, and we find an optical frequency multiplier has been involved in the local beam, with ${\alpha_1}$ > 1 as the factor. The scaling factor at the incoming beam is ${\alpha_2}$, and ${\alpha_2}$ < 1.
\begin{figure}[htbp]
\includegraphics[width=0.50\textwidth]{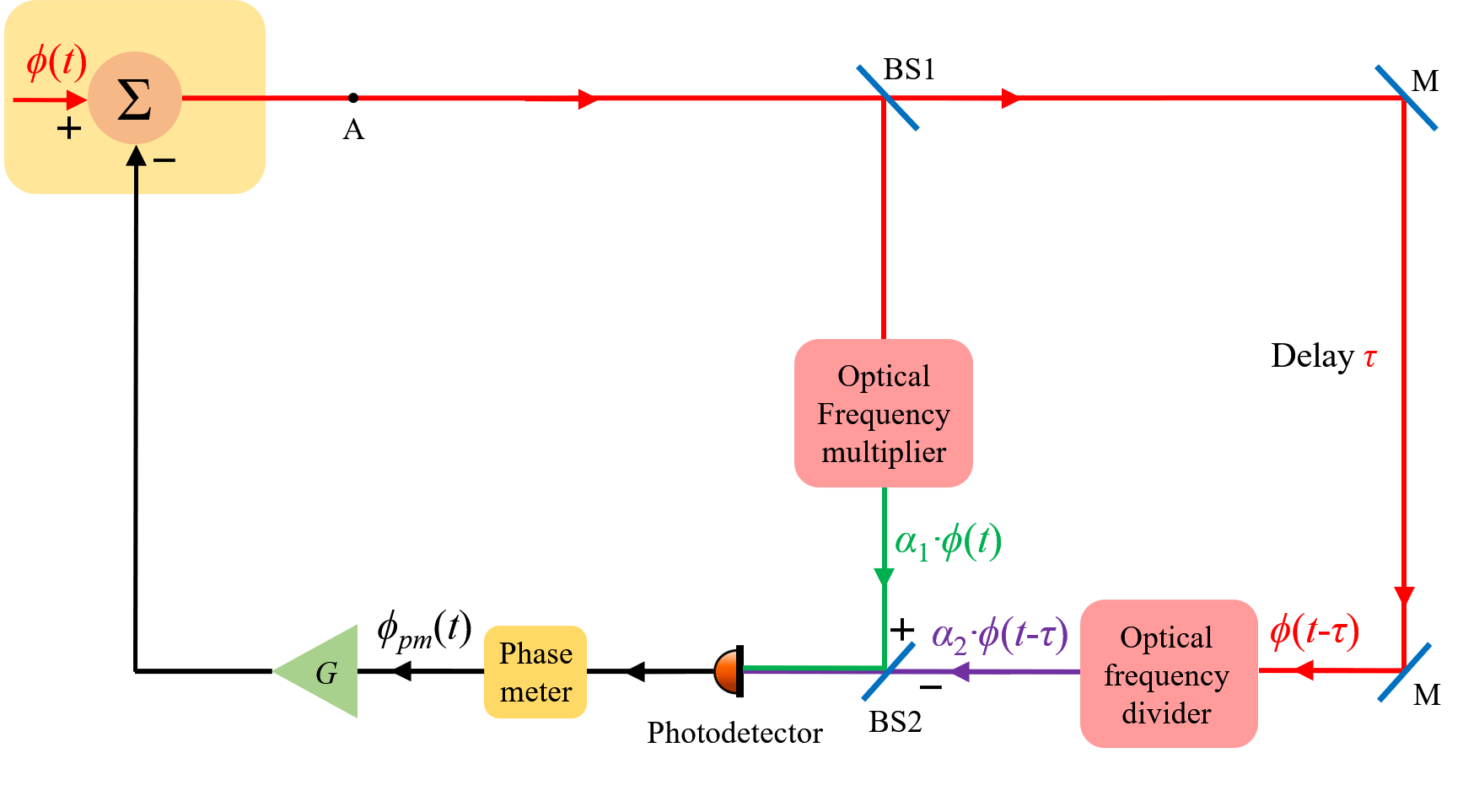}
\caption{\label{fig10} Schematic of the updated single arm locking. Both the local oscillator and the incoming beam are tuned by a certain factor.}
\end{figure}

In this case, Eq. (3) can be updated to
\begin{equation}\label{N30}
{\phi _A}\left( \omega  \right) = \frac{{\phi \left( \omega  \right)}}{{1 + G\left( \omega  \right) \cdot \left( {{\alpha _1} - {\alpha _2} \cdot {e^{ - i\omega \tau }}} \right)}}.
\end{equation}
\begin{figure}[htbp]
\includegraphics[width=0.40\textwidth]{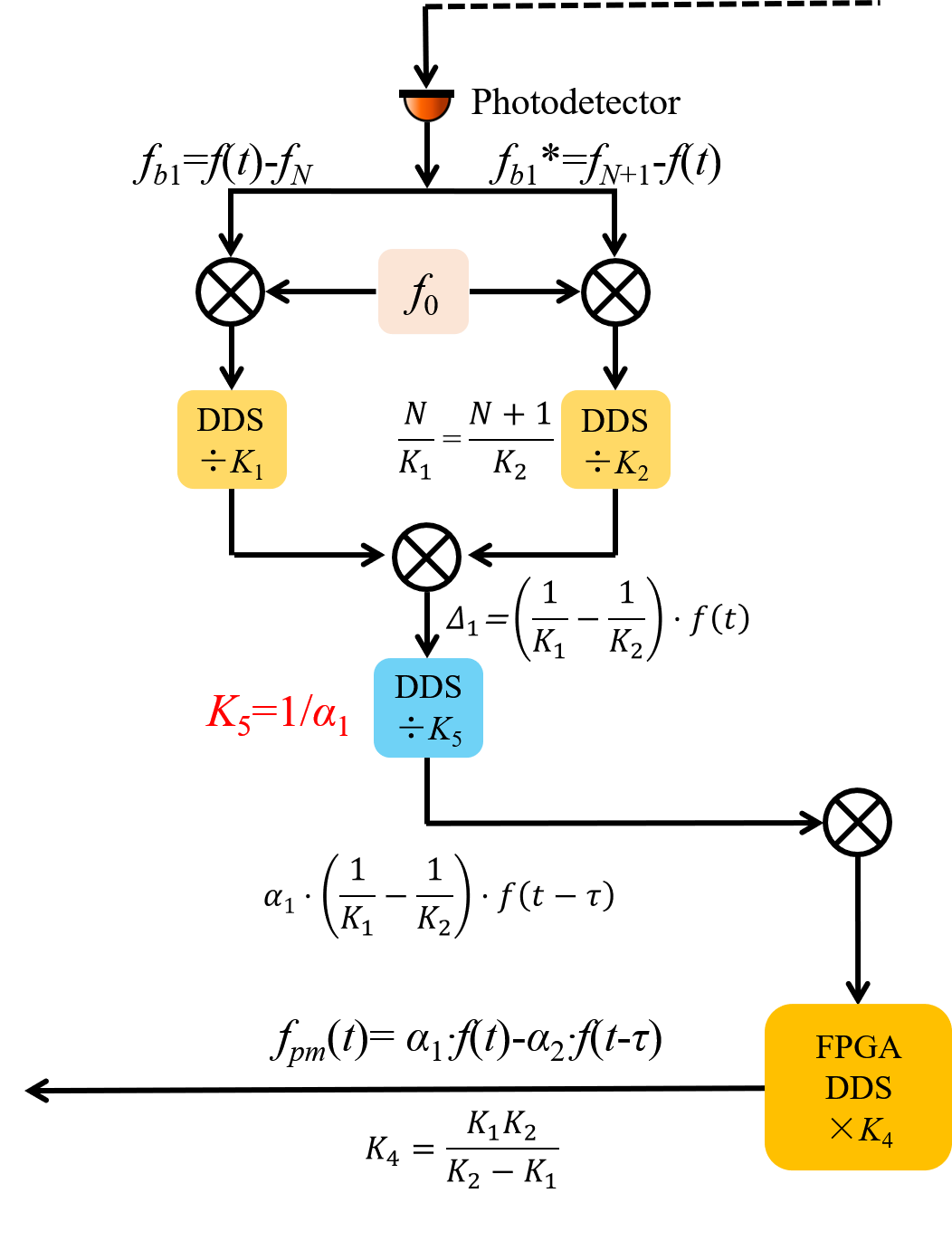}
\caption{\label{fig11} Modified local branch with an additional DDS with ${K_5}$ factor.}
\end{figure}
\begin{figure}[htbp]
\includegraphics[width=0.50\textwidth]{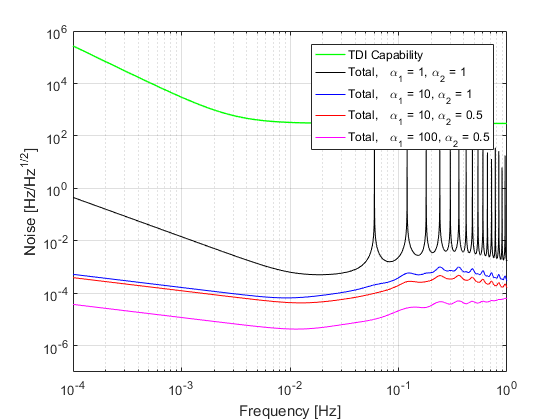}
\caption{\label{fig12} Noise floor with different configurations of (${\alpha_2}$, ${\alpha_2}$).}
\end{figure}

The arm-locking sensor is therefore ${{\alpha _1} - {\alpha _2} \cdot {e^{ - i\omega \tau }}}$, i.e., ${{\alpha _1} \cdot (1 - {\alpha _2} / {\alpha _1}{e^{ - i\omega \tau }})}$. ${{\alpha _2}/{\alpha _1}}$ <1. It is obvious that, this system is unconditionally stable. Considering the magnitude frequency response, a factor of ${\alpha _1}$ is additionally involved, compared with the sensor in Sec. 2. This can bring an additional gain in advance before the controller, which can improve the noise floor. The practical scheme shall be modified slightly. In contrast to the electrical network shown in Fig. 3, a DDS with ${K_5}$ factor should be placed in the local branch, as depicted in Fig. 11. ${K_5 = 1 / \alpha _1}$.
\begin{equation}\label{N31}
\begin{aligned}
\phi _{A,CL}(\omega) &= \frac{{{\phi _{L1}}( \omega)}}{{1 + G_1( \omega)( {{\alpha _1} - {\alpha _2} \cdot {e^{ - i\omega \tau }}} )}} \\
 & - \frac{{{G_1}( \omega )}}{{1 + {G_1}(\omega)( {{\alpha _1} - {\alpha _2} \cdot {e^{ - i\omega \tau }}})}} [ \\
 & - {\alpha _2} \cdot [ {\phi _{X13}}(\omega )( {1 + {e^{ - i\omega \tau }}}) + 2{\phi _{X31}}(\omega){e^{- i\omega {\tau _{31}}}} \\
 &  - {\phi _{S31}}(\omega  ){e^{ - i\omega {\tau _{31}}}} - \phi _{C31}(\omega){e^{- i\omega \tau _{31}}} ]\\
 & + {{\alpha _2}} \cdot {\phi _{S13}}(\omega ) + {\phi _{C13}}( \omega ) ].
\end{aligned}
\end{equation}

We examine the noise floor determined by the clock noise, the spacecraft motion, and the shot noise. (${\alpha _1}$, ${\alpha _2}$) are set to (1, 1), (10, 1), (10, 0.5), and (100, 0.5), respectively. The results are shown in Fig. 12. It is clear that, the peaks have been effectively reduced when the local oscillator or the incoming beam are tuned with factors ${\alpha _1}$ and ${\alpha _2}$. With ${\alpha _1}$ > 1, it could be valuable that the noise floor can be improved, which means that the clock noise, the noise due to the spacecraft motion, and the shot noise can be reduced in the arm locking. Nevertheless, these noise can be only kept the same in the traditional single arm locking, and amplified at the nulls. In the case of ${\alpha _1}$ = 100 and ${\alpha _2}$ = 0.5, the total noise can already be below ${10^{-4} \, \rm Hz/Hz^{1/2}}$ in the band from 0.1 mHz to 1 Hz, which is far better than the TDI capability. The performance of the noise reduction is similar with that in Sec. 4.2, other than the total noise with the free-running laser can acquire an additional reduction in advance due to the factor of ${\alpha _1}$.

\section{Conclusion}\label{section6}
In this work, we develop an updated version of single arm locking, which enables the suppression of the noise amplification at the nulls in the science band. To remove the peaks in the science band, we precisely control the laser phase noise by using laser frequency comb. The laser phase noise can be coherently scaled down into the microwave regime, which is immune to the comb frequency noises. We detailed analyze the principle of how the updated single arm locking can suppress the noise amplification at the nulls, and provide a guideline of the design of the ${\alpha}$ value. The updated single arm-locking sensor is unconditional stable when ${\alpha}$ < 1. The frequency response of the sensor shows the magnitude and the phase become more stable with decreasing the ${\alpha}$ value, which may bring advantages in the following design of the locking controller.

The analytical results show that, with the consideration of the clock noise, the spacecraft motion, and the shot noise, the peaks in the noise floor have been well suppressed, related to the factor of ${1/(1-\alpha)}$. We find that, smaller ${\alpha}$ is able to bring better noise floor. When the laser phase noise is from the free-running laser, the suppression performance can well satisfy the requirement of TDI capability, but can not achieve the noise floor owing to the limited gain. In spite of this, as long as the laser source is pre stabilized to a Mach-Zehnder interferometer or a Fabry-Perot cavity, the performance can immediately approach the noise floor, and significantly reduce the risk of TDI. Since the updated single arm locking only uses one arm of the constellation, the potential failure of the dual arm locking does not exist. This implies that, our method can serve for full year. Finally, we give a unified description of the updated single arm locking, where both the local and the incoming beams are controlled based on optical frequency comb. It could be useful that, the noise floor can be improved as well. Arm locking will be a valuable solution of the laser phase noise for the space-borne GW detector in future. No additional hardware is required for the realization of arm locking, but the possible risk of TDI can be greatly reduced. Recently, the space-borne frequency combs have been reported, showing the powerful capability in the space mission [30]. Meanwhile, various comb sources are developing at a rapid pace [31,32], which can be the candidates for the future GW detectors. Our work provides a simplified and ease-to-use method for the arm locking, and can be used in the operation in orbit.\\

\section*{ACKNOWLEDGMENTS}
This work is supported by National Key R\rm \&D Program of China (Grant No. 2020YFC2200500), the National Natural Science Foundation of China (Grant Nos. 11575160, 11805074, and 11925503), Guangdong Major Project of Basic and Applied Basic Research (Grant No. 2019B030302001), and the Fundamental Research Funds for the Central Universities, HUST: 2172019kfyRCPY029.

\end{document}